\documentclass[conference,usenatbib]{basi}
\usepackage[british]{babel}
\usepackage[varg]{txfonts}
\usepackage{rotating}
\usepackage{dcolumn}
\begin{document}
\title{Homogenized effective temperatures from stellar libraries}
\author[V.Malyuto]%
       {V.~Malyuto
       \\
       $$Tartu Observatory, 61602 Tartumaa, T\~oravere, Estonia}

\pubyear{2012}
\volume{00}
\pagerange{\pageref{firstpage}--\pageref{lastpage}}
\editors {P. Prugniel and H.P. Singh (Editors)}
\date{Received 26 January 2012 / Accepted 7 March 2012}
\maketitle
\label{firstpage}

\begin{abstract}
External errors of effective temperatures of stars for selected libraries are estimated
from data intercomparisons. It is found that the
obtained errors are mainly in a good correspondence with the published data. 
The results may be used to homogenize the effective temperatures by averaging the data 
(with the weights inversely 
proportional to the squared errors) from independent sources.
\end{abstract}


\section{Introduction}
Numerous spectral and photometric stellar catalogues of the basic atmospheric parameters 
($T_{\rm eff}$, $\log g$, [Fe/H]) are widely used 
for decoding the structure, evolutionary stage and chemical enrichment history of the 
Galaxy. The rapidly growing number of such catalogues has
imposed a need for refining procedures of merging these  stellar data
into a single  homogenized catalogue.
In our recent paper \citep{2011BaltA..20...91M}, the technique has been applied where 
some homogeneous samples from selected catalogues of
the $T_{\rm eff}$ values were treated by combining them in triples, quadruples, quintuples and pairs 
for the stars in common to determine their external errors 
from data intercomparisons. The $T_{\rm eff}$ values are then averaged (with the weights inversely 
proportional to the squared errors) to produce an extended mean homogenized catalogue. 
A somewhat different procedure of homogenization was used by \citet{2007MNRAS.374..664C} 
where only pairs of stars were treated and heterogeneity of the published errors (or other quality data) of the parameters  
inside every used sample was not taken into account. 

The most important and detailed information for stars is available in stellar spectral
libraries with medium to high resolutions and good coverages of the 
Herzsprung-Russell diagram and metallicity range. Some libraries also contain the results of an
efficient parametrization of $T_{\rm eff}$, $\log g$, [Fe/H]. Few libraries are considered here whose published $T_{\rm eff}$
are compared with the  $T_{\rm eff}$ from an independent extensive homogeneous sample 
with the external ${\sigma T_{\rm eff}}$ taken from our previous analysis \citep{2011BaltA..20...91M}.
These comparisons allow to estimate the ${\sigma T_{\rm eff}}$ for the 
considered spectral libraries, and to use the results for homogenizing the $T_{\rm eff}$ values.

\section{Results}
To estimate ${\sigma T_{\rm eff}}$ from data intercomparisons,  \citet{2011BaltA..20...91M} have selected a set of homogeneous samples. 
Among them there is the most populated 
and homogeneous sample (S.1) from the photometric catalogue of  \citet{2006A&A...450..735M} with the published $T_{\rm eff}$=5200-6700 K, 
${\sigma T_{\rm eff}}$=40-60 K and [Fe/H] $>$ -1.1.  
For comparison with S.1, the  
following samples (S.2-S.5) with the stars in common were selected from the appropriate independent spectral libraries:

\vspace{-4mm}
S.2 \citep{2007yCat.3251....0P} -- with the $T_{\rm eff}$ compiled from the literature for ELODIE library.

\vspace{-4mm}
S.3 \citep{2011A&A...525A..71W} -- with the $T_{\rm eff}$ determined the use of the ULySS package from CFLIB library.

\vspace{-4mm}
S.4 \citep{2011A&A...531A.165P} -- with the $T_{\rm eff}$ determined the use of the ULySS package from MILES library.

\vspace{-4mm}
S.5 \citep{2007MNRAS.374..664C} -- with the $T_{\rm eff}$ compiled from the literature for MILES library.

\begin{figure}
%
\centerline{\includegraphics[width=11cm,angle=270]{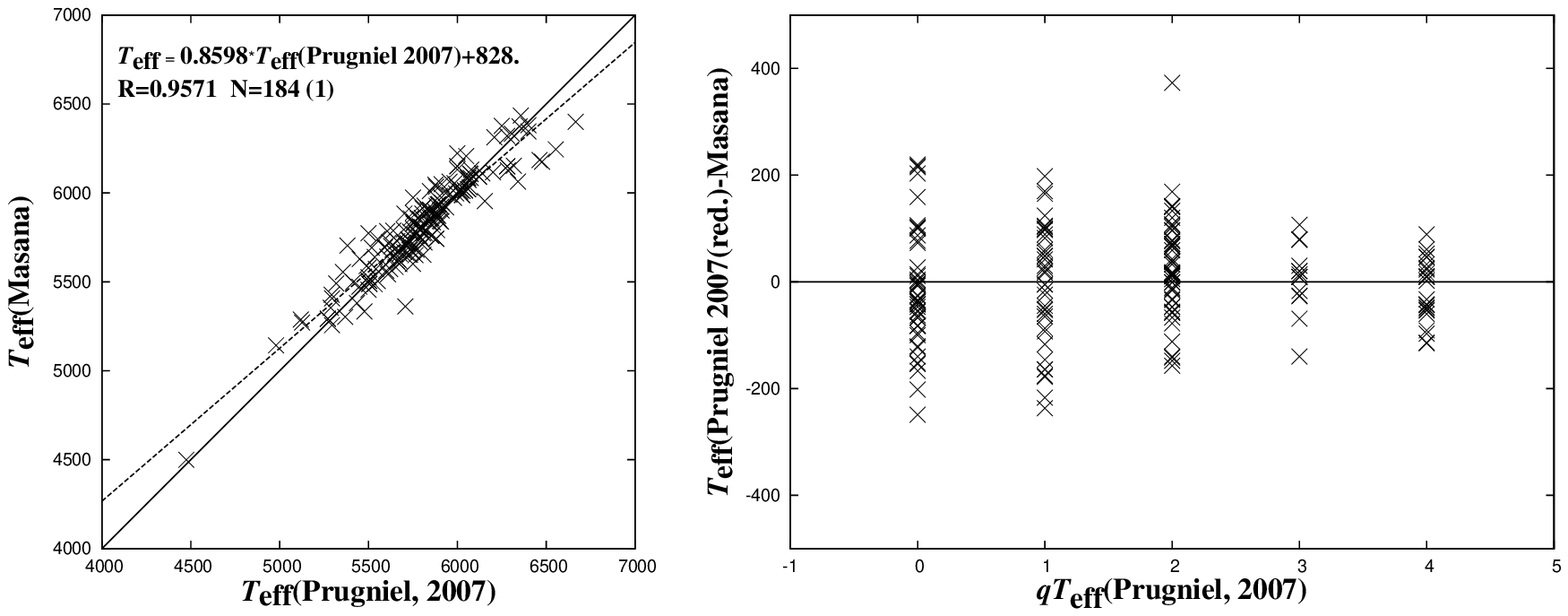}}
\vspace{-6.5cm}
\centerline{\includegraphics[width=11cm,angle=270]{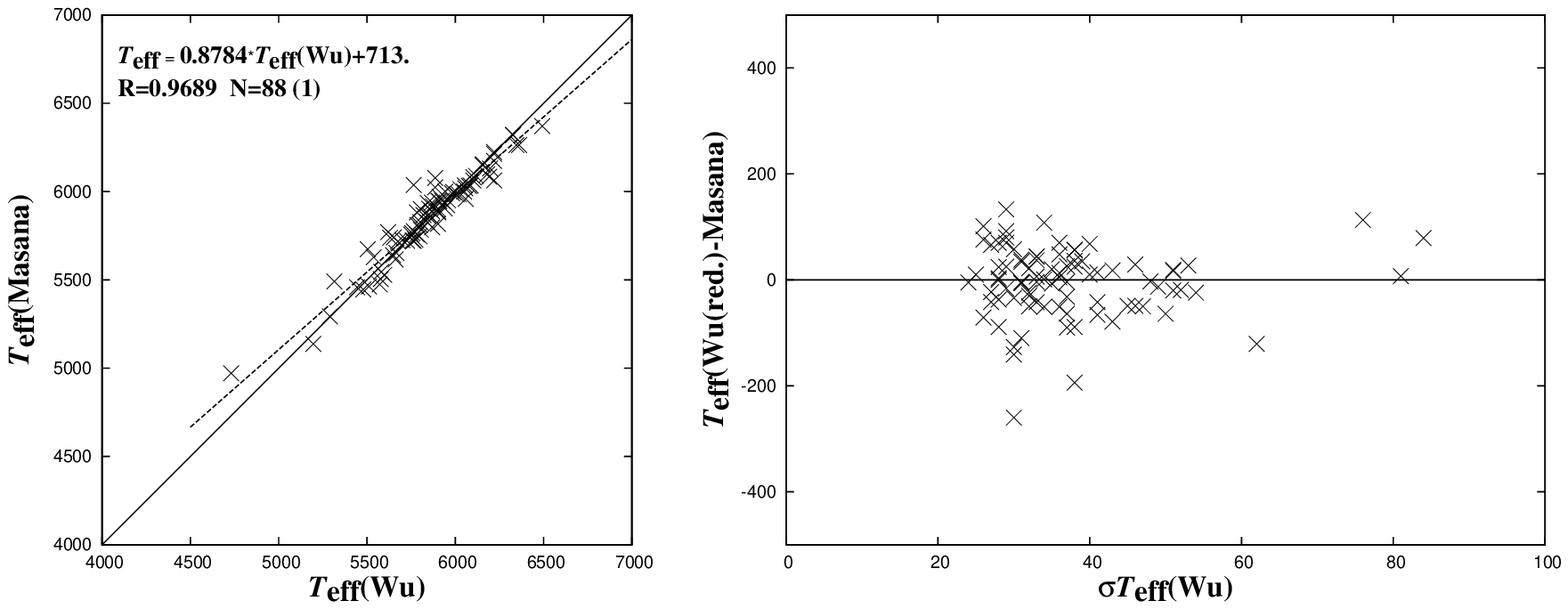}}
\vspace{-6.5cm}
\centerline{\includegraphics[width=11cm,angle=270]{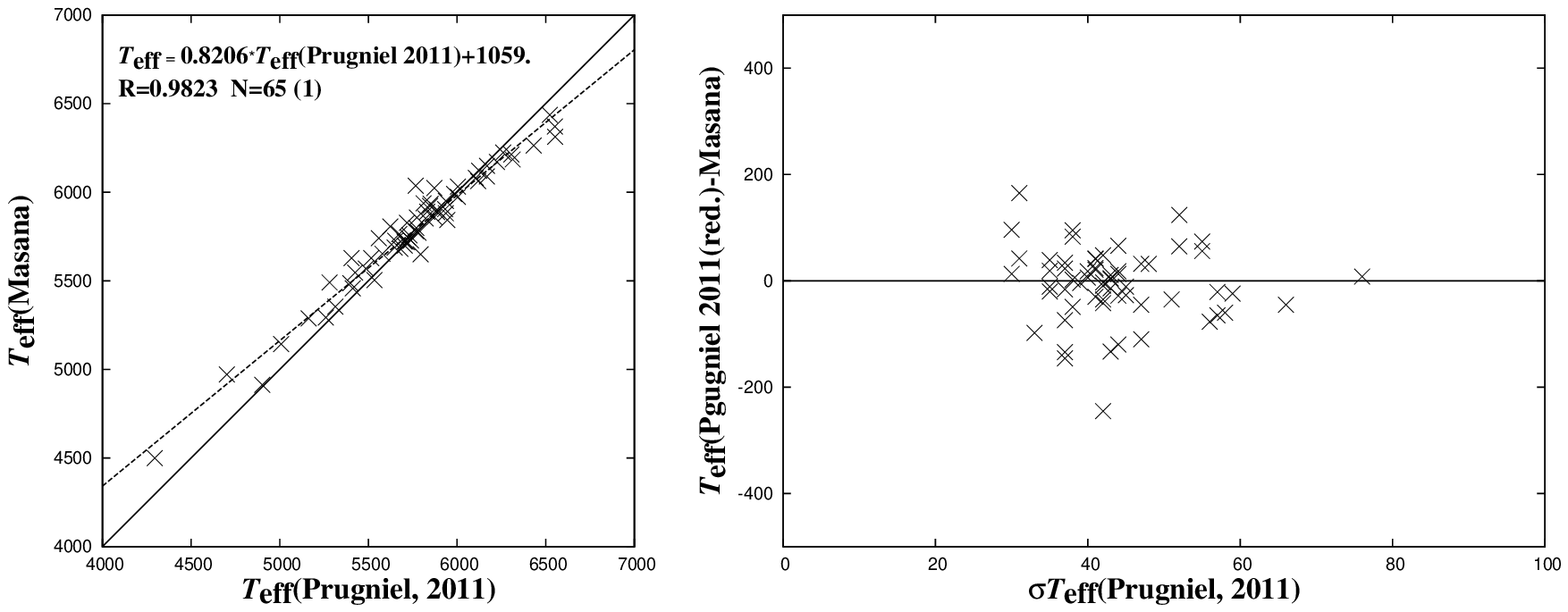}}
\vspace{-6.5cm}
\begin{tabular}{p{4.5cm}cp{6cm}}
\raisebox{-\height}{\includegraphics[width=11cm,angle=270,clip=2cm]{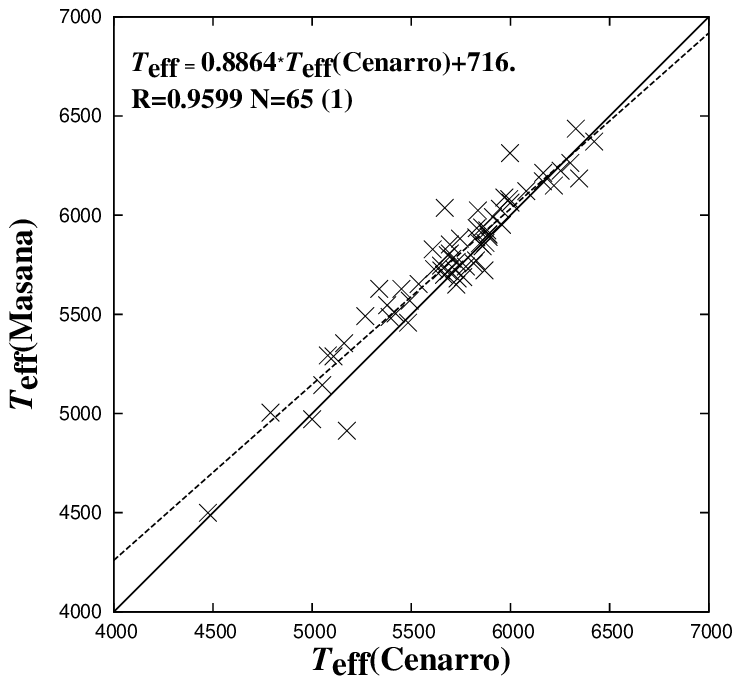}} & \quad &
\caption{Comparisons of some published data from S.1-S.5. The reductions (broken lines)
of all $T_{\rm eff}$ to the system of \citet{2006A&A...450..735M} with their correlation coefficients $R$ and
the numbers of stars in common $N$, are presented
on the left.
The numbers of stars in common rejected from our calculations according to the 
$3\sigma$ rule,
are given in brackets. The reduced data for $T_{\rm eff}$ are presented  on the right.\label{f:narrow}}
\end{tabular}
\end{figure}

Some comparisons of the data are presented in Fig.~\ref{f:narrow}. The $T_{\rm eff}$ from these libraries were reduced to the system of 
\citet{2006A&A...450..735M} with the
equations given on the left in this Figure. The $T_{\rm eff}$ differences are presented on the right in the same Figure versus
the published quality data 
$qT_{\rm eff}$ from S.2 \citep{2007yCat.3251....0P} or the published ${\sigma T_{\rm eff}}$ from S.3 \citep{2011A&A...525A..71W}
and 
from S.4 \citep{2011A&A...531A.165P}, respectively. We see that the scatter depends on the $qT_{\rm eff}$ for S.2 \citep{2007yCat.3251....0P},
as expected. To deal with 
homogeneous data, we divided S.2 into three subsamples: S.2A with the $qT_{\rm eff}$=3-4, S.2B
with the $qT_{\rm eff}$=2 and S.2C with the $qT_{\rm eff}$=0-1 (the latter data are of the lowest quality). The  scatters do not 
depend on the ${\sigma T_{\rm eff}}$ for S.3 \citep{2011A&A...525A..71W} and for S.4 \citep{2011A&A...531A.165P}
in this Figure, and there is no information on the quality data or ${\sigma T_{\rm eff}}$ for S.5 \citep{2007MNRAS.374..664C}. 
\begin{table}[]
\caption{External ${\sigma T_{\rm eff}}$ for S.2-S.5 obtained from the comparison with
S.1. Some published data and the numbers of stars in common $N$ are given at the end of this Table.
\label{f:errors}}
\begin{center}
\noindent
\begin{tabular}{ccccccc}
\hline
Pair \hfill & S.2A \hfill & S.2B \hfill & S.2C \hfill &S.3 \hfill &S.4 \hfill &S.5
\\
\hline
S.1, S.2A & 33 & -- & -- & -- & -- & --\\
S.1, S.2B & -- & 64 & -- & -- & -- & --\\
S.1, S.2C & -- & -- & 98 & -- & -- & --\\
S.1, S.3  & -- & -- & -- & 35 & -- & --\\
S.1, S.4  & -- & -- & -- & -- & 38 & --\\
S.1, S.5  & -- & -- & -- & -- & -- & 80\\
\hline
Published ${qT_{\rm eff}}$& 3-4  & 2
& 0-1   &--&--&--\\
Mean published ${\sigma T_{\rm eff}}$& --   &~-
& --   &37$\pm 11$&44$\pm 9$&--\\
\hline
$N$&42&51&88&86&60&60\\
\hline
\noalign{\vskip2mm}
\end{tabular}
\end{center}
\noindent
\end{table}

The variances for $T_{\rm eff}$ differences between S.1, on the one hand, and S.2A, S.2B, S.2C, S.3, S.4 and S.5, on the other hand, 
were calculated, respectively. Then the external ${\sigma T_{\rm eff}}$ for the appropriate samples
were estimated from the variances through  extracting the known ${\sigma T_{\rm eff}}$=49 K for S.1, 
the latter value  was taken from \citet{2011BaltA..20...91M}.
The results are presented 
in Table~\ref{f:errors}.
There is a nice correspondence between the external ${\sigma T_{\rm eff}}$ and the published $qT_{\rm eff}$ for 
S.2A, S.2B, S.2C, respectively. For S.3 and S.4, the obtained external ${\sigma T_{\rm eff}}$ are very simular to the
mean published  ${\sigma T_{\rm eff}}$, respectively. Therefore, the published ${\sigma T_{\rm eff}}$ should be reliable. 

However, the external ${\sigma T_{\rm eff}}$  for S.5 is significantly larger then for S.4 
(80 K and 38 K, respectively).  
The difference may occur because in the compilation for MILES library
\citet{2007MNRAS.374..664C} used the procedure of data homogenization where 
~~heterogeneity of the published errors (or other quality data) of the parameters inside every used sample has not 
been taken into account. 
Our comparisons of $T_{\rm eff}$  for some catalogues   
\citep{2011BaltA..20...91M} have shown that 
the use of samples
where the published  ${\sigma T_{\rm eff}}$ are within some selected intervals really helps us deal with
more homogeneous data of $T_{\rm eff}$. 
Therefore, such errors should be involved in homogenization process.
\section{Conclusions}
The technique of estimating the errors of catalogues from data intercomparisons has been applied to 
some samples of the published $T_{\rm eff}$ for spectral libraries. 
The results will be used for producing an extended mean homogenized catalogue of  $T_{\rm eff}$.
The same approach may be applied also for the treatment of other kinds of data.
The stars with the most reliable parameters  may then serve as
templates in classification. Support from the Estonian Science Foundation (grant No. 7765) is
acknowledged.

\label{lastpage}
\end{document}